\documentclass[11pt]{article}
\pdfoutput=1
\usepackage{fullpage}
\usepackage{xcolor}
\usepackage{cite}
\usepackage{graphicx}
\usepackage{tensor}
\usepackage{amsmath,bm}
\usepackage{amssymb}
\usepackage{subfig}
\usepackage{tikz}
\usepackage[utf8x]{inputenc} 
\title{}
\date{}

\def\beq{\begin{equation}}
\def\eeq{\end{equation}}
\allowdisplaybreaks

\begin{document}
\bibliographystyle{utphys}

% Commands
\newcommand{\hel}{\eta} % Helicity label
\renewcommand{\d}{\mathrm{d}}
\newcommand{\dd}{\hat{\mathrm{d}}}
\newcommand{\del}{\hat{\delta}}
\newcommand{\ket}[1]{| #1 \rangle}
\newcommand{\bra}[1]{\langle #1 |}

\newcommand{\be}{\begin{equation}}
\newcommand{\ee}{\end{equation}}
\newcommand\n[1]{\textcolor{red}{(#1)}} %in-text notes
\newcommand{\diff}{\mathop{}\!\mathrm{d}}
\newcommand{\lb}{\left}
\newcommand{\rb}{\right}
\newcommand{\f}{\frac}
\newcommand{\pd}{\partial}
\newcommand{\tr}{\text{tr}}
\newcommand{\fdiff}{\mathcal{D}}
\newcommand{\im}{\text{im}}
\let\caron\v
\renewcommand{\v}{\mathbf}
\newcommand{\T}{\tensor}
\newcommand{\R}{\mathbb{R}}
\newcommand{\C}{\mathbb{C}}
\newcommand{\Z}{\mathbb{Z}}
\newcommand{\msbar}{\ensuremath{\overline{\text{MS}}}}
\newcommand{\DIS}{\ensuremath{\text{DIS}}}
\newcommand{\abar}{\ensuremath{\bar{\alpha}_S}}
\newcommand{\bb}{\ensuremath{\bar{\beta}_0}}
\newcommand{\rc}{\ensuremath{r_{\text{cut}}}}
\newcommand{\Nd}{\ensuremath{N_{\text{d.o.f.}}}}
\newcommand{\red}[1]{{\color{red} #1}}

\newcommand{\Ad}{\dot{A}}
\newcommand{\Bd}{\dot{B}}
\newcommand{\Cd}{\dot{C}}
\newcommand{\Dd}{\dot{D}}
\newcommand{\Ed}{\dot{E}}
\newcommand{\Fd}{\dot{F}}
\newcommand{\depsilon}{\epsilon}
\newcommand{\dsigma}{\bar{\sigma}}

\newcommand{\bphi}{\phi} % rescaled dilaton
\newcommand{\bB}{B} % rescaled B field
\newcommand{\bH}{H} % rescaled H
\newcommand{\bsigma}{\sigma} % rescaled axion
\newcommand{\charge}{\tilde{c}} % Constant for dilaton and axion classical fields
\newcommand{\ampA}{\mathcal{A}} % amplitude A: 3pt gauge amplitude
\newcommand{\ampM}{\mathcal{M}} % amplitude M: 3pt gravity amplitude
\newcommand{\td}[1]{%
  \medskip
  \par\noindent
  \fcolorbox{black!30}{yellow!20}{%
    \parbox{\dimexpr\linewidth-2\fboxsep-2\fboxrule}{\scriptsize #1}%
  }%
  \medskip
}
\renewcommand{\[}{\begin{equation}\begin{aligned}}
\renewcommand{\]}{\end{aligned}\end{equation}}

\titlepage
%\begin{flushright}
%QMUL-PH-22-??\\
%\end{flushright}

\vspace*{0.5cm}

\begin{center}
{\bf \Large The double copy as a doppelg\"{a}nger}

\vspace*{1cm} 
\textsc{Nathan Moynihan\footnote{n.moynihan@qmul.ac.uk}, Michael L. Reichenberg Ashby\footnote{m.l.reichenbergashby@qmul.ac.uk}
  and Chris D. White\footnote{christopher.white@qmul.ac.uk}} \\

\vspace*{0.5cm} Centre for Theoretical Physics, School of Physical and
Chemical Sciences, \\ Queen Mary University of London, 327 Mile End
Road, London E1 4NS, UK\\

\end{center}

\vspace*{0.5cm}

\begin{abstract}
The double copy relates scattering amplitudes and classical solutions
in non-abelian gauge theories and gravity. As such, it is
usually expressed in the conventional second-order formalisms
in both theories corresponding to standard Yang-Mills theory, and the Einstein--Hilbert action in
General Relativity. In this paper, we instead consider alternative
formulations of gravity, which are known to terminate at finite order
in the coupling at Lagrangian level. We focus in particular on the Chern--Simons--Witten (CSW) formulation in 2+1 dimensions, and argue that the double copy then becomes a doppelg\"{a}nger relationship between gauge theory and gravity, allowing straightforward replacement of generators and structure constants in both theories. We show how explicit (multiple) static point-source solutions can be mapped in the two approaches, and use the CSW formalism to examine when double copies are expected to be possible, and when not. In addition, we present an explicit double copy between the Wong equations for colour charges, and the Mathisson-Papapetrou-Dixon equations for spinning particles, that extends also to higher dimensions.
\end{abstract}

\vspace*{0.5cm}

\section{Introduction}
\label{sec:intro}

The double copy is a by now well-established relationship between
quantities in gauge, gravity and related theories. Inspired by
previous work in string theory~\cite{Kawai:1985xq}, its original
incarnation applied to scattering amplitudes in perturbation
theory~\cite{Bern:2008qj,Bern:2010ue,Bern:2010yg}, before it was
extended to classical
solutions~\cite{Monteiro:2014cda,Luna:2015paa,Ridgway:2015fdl,Bahjat-Abbas:2017htu,Carrillo-Gonzalez:2017iyj,CarrilloGonzalez:2019gof,Bah:2019sda,Alkac:2021seh,Alkac:2022tvc,Luna:2018dpt,Sabharwal:2019ngs,Alawadhi:2020jrv,Godazgar:2020zbv,White:2020sfn,Chacon:2020fmr,Chacon:2021wbr,Chacon:2021hfe,Chacon:2021lox,Dempsey:2022sls,Emond:2021lfy,Easson:2022zoh,Chawla:2022ogv,Han:2022mze,Armstrong-Williams:2022apo,Han:2022ubu,Elor:2020nqe,Farnsworth:2021wvs,Anastasiou:2014qba,LopesCardoso:2018xes,Anastasiou:2018rdx,Luna:2020adi,Borsten:2020xbt,Borsten:2020zgj,Goldberger:2017frp,Goldberger:2017vcg,Goldberger:2017ogt,Goldberger:2019xef,Goldberger:2016iau,Prabhu:2020avf,Luna:2016hge,Luna:2017dtq,Cheung:2016prv,Cheung:2021zvb,Cheung:2022vnd,Cheung:2022mix,Chawla:2024mse,Keeler:2024bdt,Chawla:2023bsu,Easson:2020esh,Armstrong-Williams:2024bog,Armstrong-Williams:2023ssz,Farnsworth:2023mff,Emond:2025nxa}. Potential
non-perturbative aspects have been addressed in
ref.~\cite{Monteiro:2011pc,Borsten:2021hua,Alawadhi:2019urr,Banerjee:2019saj,Huang:2019cja,Berman:2018hwd,Alfonsi:2020lub,Alawadhi:2021uie,White:2016jzc,DeSmet:2017rve,Bahjat-Abbas:2018vgo,Armstrong-Williams:2025spu,Cheung:2022mix,Moynihan:2021rwh,Borsten:2022vtg},
and recent pedagogical reviews may be found in
ref.~\cite{Borsten:2020bgv,Bern:2019prr,Adamo:2022dcm,Bern:2022wqg,White:2021gvv,White:2024pve}. Despite
the above progress, a full understanding of the scope and remit of the
double copy remains lacking, due mainly to a failure to understand its
operation at the level of Lagrangians or, equivalently, equations of
motion. This is chiefly due to the fact that a certain relationship
between the colour and kinematic degrees of freedom~\cite{Bern:2008qj}
-- known as BCJ duality -- must be made manifest in order to
double-copy non-linear amplitudes or solutions. This can typically be
achieved only order-by-order in perturbation theory, resulting in a
gauge theory Lagrangian that involves increasingly complicated
higher-order interactions (see e.g. ref.~\cite{Bern:2010yg}). Upon
performing the double copy, the structure of such interactions is such
as to reproduce the all-order form of the conventional
Einstein--Hilbert gravitational action. Although the latter looks
simple in that it involves a single power of the Ricci scalar, one
finds an infinite tower of higher-order graviton interactions upon
expanding the action (together with the covariant spacetime volume
measure) in terms of the graviton field.

As is well-known, the Einstein--Hilbert formulation is not the only way
to write down General Relativity and related theories. Many
alternative approaches exist, in which the role of the graviton is
replaced by dependence on other quantities of geometric interest, such
as the vierbein and / or spin connection\footnote{A detailed and
highly pedagogical review of alternative formulations of General
Relativity may be found in ref.~\cite{Krasnov:2020lku}.}. The action
then terminates at finite order in the gravitational coupling
constant, and the question then naturally arises of what the double
copy looks like in such formulations. Is it still a ``double copy'' in
the usual sense, or are other replacement rules or structures at play?
Answering such questions may prove fruitful in elucidating the origins
and scope of the double copy, as well as the mysterious {\it kinematic
  algebras} that are known to be a consequence of BCJ duality \cite{Reiterer:2019dys,Ben-Shahar:2024dju,Armstrong-Williams:2022apo,Mizera:2019blq,Borsten:2022vtg,Armstrong-Williams:2024icu,Diaz-Jaramillo:2025gxw}. Both
the double copy and BCJ duality strongly suggest that our traditional
language for describing field theories is hiding a common underlying
structure. The use of alternative formulations for describing gravity
could help uncover this structure, and in any case having multiple
languages to express the same physical idea -- where the ``right''
language depends upon the question being asked -- is often a fruitful
way to proceed in physics.

The main theme of our paper will be to adopt formulations of gravity
in which the action and / or equations of motion look identical in
form to those of (non-abelian) gauge theory. In such formulations, the
double copy ceases to look like a ``doubling'', and we instead
(following ref.~\cite{Cheung:2021zvb}) refer to the gravity theory as
the doppelg\"{a}nger of the gauge theory. To go from one theory to the
other, one simply reinterprets a colour index in terms of an internal
index in gravity theory, which is typically associated with a
generator of the Lorentz or Poincar\'{e} algebra. Upon translating to the usual
Einstein--Hilbert formulation of gravity, the conventional double copy
emerges. For the cases we study, we may therefore regard the
doppelg\"{a}nger relationship as an underlying explanation for the
double copy.

We note that our work is in a similar spirit to -- and indeed inspired by -- the so-called {\it covariant colour-kinematics duality} of ref.~\cite{Cheung:2021zvb}. This also considered manifest doppelg\"{a}nger relationships at the level of the equations of motion, and such that types of index were reinterpreted in order to proceed from one theory to another. Our results are distinct from those in ref.~\cite{Cheung:2021zvb}, however, in that we consider different formalisms of gravity, and also show how explicit solutions can be mapped between theories. 

We will begin in the following section by reviewing the various formulations of gravity that we will use throughout the paper, in particular the well-known Einstein-Cartan formalism for General Relativity. Unlike the conventional Einstein--Hilbert description of gravity in terms of the metric tensor, the Einstein-Cartan approach instead introduces two dynamical quantities known as the {\it vielbein} (in $d$ dimensions) and the {\it spin connection}, from which the conventional metric and curvature tensors can be obtained. Advantages of this formalism include the fact that the Lagrangian is manifestly polynomial, and also that one may furnish the theory with a suggestive geometric interpretation that mimics the structure of (non-abelian) gauge theories in terms of fibre bundles. In 2+1 dimensions, this idea can be taken further, such that gravity can be entirely written as a gauge theory of the Poincar\'{e} group. This is the so-called Chern--Simons--Witten (CSW) formulation of three-dimensional gravity~\cite{Witten:1988hc}, and we also discuss its coupling to matter particles. We will argue in detail that in this language, the double copy between gauge theory and gravity can be reinterpreted as a simple replacement of colour indices with gravitational internal indices (corresponding to the gauged Poincar\'{e} symmetry). Furthermore, coupling to matter introduces degrees of freedom living on the worldline in both the gauge and gravity cases. These degrees of freedom can themselves be put in a similar form in both the gauge and gravity theory, again such that internal indices get reinterpreted in going from one theory to the other.

Our analysis could be taken as an underlying explanation for the double copy in 2+1 dimensions, with the existence of a gauge-theoretic formulation of gravity naturally implying a double-copy structure when translated into the conventional Einstein--Hilbert description. Indeed, we will show how the conventional double copy of Wilson lines~\cite{Alfonsi:2020lub} emerges from the Chern--Simons--Witten theory. This is already interesting, given that Wilson lines capture all-order properties of perturbative scattering amplitudes. More importantly, though, our ideas can be used to resolve a key open question regarding the scope of the double copy itself: given an exact solution in a gauge theory, is it always possible to convert this into an exact solution in gravity? We will show precisely how this mapping works for pointlike sources in CSW theory, including how indices in the CSW gauge field get reinterpreted in terms of colour degrees of freedom. Crucial to this analysis is that one can identify a subset of generators of the Poincar\'{e} group in gravity that get mapped to a given subalgebra of the gauge group in the single copy theory. It is unlikely such a mapping exists for arbitrary solutions, thereby diminishing previous hopes that one might be able to double copy arbitrary non-abelian solutions.

The structure of our paper is as follows. In section~\ref{sec:review}, we review relevant formulations of gravity needed for the rest of the paper. In section~\ref{sec:2+1}, we study the double copy in 2+1 dimensions in detail, showing how this can be reinterpreted as a set of straightforward replacements between doppelg\"{a}nger theories. We also show how classical topological solutions can be explicitly mapped, and how this can fail. Finally, we discuss our results and conclude in section~\ref{sec:conclude}.

\section{Alternative formulations of General Relativity}
\label{sec:review}

Einstein's formulation of General Relativity in terms of a metric tensor is not the only one. Various formulations of the theory exist by now, which use a variety of mathematical languages. Different ways of thinking about gravity can offer calculational advantages, but also conceptual advantages, particularly in relating gravitational physics to well-studied geometric structures such as fibre bundles. Here, we review salient details of formulations needed for for the rest of the paper, where our presentation follows that of ref.~\cite{Krasnov:2020lku} in places.

\subsection{Einstein-Cartan Theory}
\label{sec:EC}

Perhaps the most well-known alternative formulation of gravity is the {\it Einstein-Cartan formalism}. As stressed in ref.~\cite{Krasnov:2020lku}, this can itself be introduced in different ways, and we here adopt the geometric point of view adopted in that reference. Given a (spacetime) manifold $M$, we may construct a vector bundle $E\rightarrow M$, namely a space $E$ which is locally isomorphic to $M\times V$ for some vector space $V$. For Einstein-Cartan theory, one requires the vector space $V$ to be equipped with an inner product, and for the bundle $E$ to be associated to the principal frame bundle such that it is isomorphic to the tangent bundle $TM$. The latter is achieved by introducing $d$ linearly independent one-forms
\begin{equation}
  e^a={e^a}_\mu dx^\mu,
  \label{eadef}
\end{equation}
known as the {\it vielbein} in $d$ spacetime dimensions. Here the vector on the left-hand side lives in the internal space $V$, and the coefficients on the right-hand side can be used to convert vectors $v=v^\mu\partial_\mu$ from the tangent space of $M$ to the internal space via
\begin{equation}
  e^a(v)={e^a}_\mu v^\mu.
  \label{eadef2}
\end{equation}
Let $\langle \cdot ,\cdot \rangle$ denote the inner product between two vectors in the internal space, which we may write in components as
\begin{equation}
  \langle {\bf v}_1,{\bf v}_2\rangle=\eta_{ab}v_1^a v_2^b.
  \label{metricdef}
\end{equation}
One may then pullback this metric to obtain the corresponding metric on the spacetime manifold $M$. That is, if $u$ and $v$ denote vectors in the tangent space of $M$, then
\begin{equation}
  g(u,v)=\langle e(u),e(v)\rangle,\quad g_{\mu\nu}={e^a}_\mu {e^b}_\nu \eta_{ab}.
\label{metricdef2}
\end{equation}
Note that the choice of vielbein is not unique, but may be subjected to local transformations that preserve the internal space metric. Choosing this metric to be the Minkowski metric, these transformations will be Lorentz:
\begin{equation}
  e^a={ \Lambda^a}_b e^b,\quad \eta_{ab}={\Lambda^c}_a{\Lambda^d}_b \eta_{cd}.
  \label{Lorentz}
\end{equation}
Furthermore, in order that the spacetime metric be non-degenerate, it is required that the vielbein have a suitable inverse ${e^\mu}_a$ such that
\begin{equation}
    {e^a}_\mu {e^\mu}_b = \delta^a_b,\quad {e^\mu}_a {e^a}_\nu = \delta^\mu_\nu
\end{equation}
We may then define a connection on the bundle $E$, which in turn is used to form covariant derivatives of vector fields:
\begin{equation}
  D V^a=dV^a+{\omega^a}_b\wedge V^b.
  \label{covD}
\end{equation}
The one-form ${\omega^a}_b$ is known as the {\it spin connection}, and is required to be {\it metric-compatible} meaning vanishing covariant derivative of the internal space metric:
\begin{equation}
  D\eta^{ab}={\omega^a}_c \eta^{cb}+{\omega^b}_c\eta^{ac}=0.
  \label{metcomp}
\end{equation}
The conventional covariant derivative $\nabla$ that acts on $TM$ can be obtained by the pullback relation
\begin{equation}
  e^a_\mu \nabla u^\mu = D u^a,
  \label{nabladef}
\end{equation}
from which one may obtain the following relation between the spin connection and Levi-Civita connection $\Gamma^\nu_{\mu\lambda}$:
\begin{equation}
  \Gamma^\nu_{\mu\lambda}={e^\nu}_a \partial_\mu {e_\lambda}^a
  +{e^\nu}_a{e_\lambda}^b{{\omega_\mu}^a}_b.
  \label{Gammadef}
\end{equation}
This is already to sufficient to tell us that, in contrast the usual Einstein equations that depend only upon the spacetime metric $g_{\mu\nu}$, the Einstein-Cartan formalism is instead phrased in terms of two dynamical quantities, namely the vielbein and spin connection. In order to state the equations of motion, one may introduce the {\it torsion}, or Lorentz-covariant derivative of the vielbein:
\begin{equation}
  T^a=D e^a=\d e^a+{\omega^a}_b \wedge e^b\equiv {T^a}_{\mu\nu}dx^\mu\wedge
  dx^\nu.
  \label{Tadef}
\end{equation}
Converting the internal index to a spacetime one and using eq.~(\ref{Gammadef}) yields
\begin{equation}
  {T^\lambda}_{\mu\nu}={e^\lambda}_a {T^a}_{\mu\nu}=
  \Gamma^\lambda_{\mu\nu}-\Gamma^\lambda_{\nu\mu},
  \label{torsion}
\end{equation}
such that eq.~(\ref{Tadef}) is indeed related to the usual torsion in terms of the Christoffel symbol. Vanishing of the torsion is then expressed by the equation
\begin{equation}
  \d e^a+{\omega^a}_b \wedge e^b=0.
  \label{EC1}
\end{equation}
We may also introduce an analogue of the Riemann curvature, by considering the commutator of two covariant derivatives on a vector in the internal space. In components:
\begin{equation}
  [D_\mu,D_\nu]u^a\equiv {{R^a}_b}_{\mu\nu}u^b,
  \label{curv1}
\end{equation}
which defines a two-form
\begin{equation}
  {R^a}_b=\d{\omega^a}_b+{\omega^a}_c\wedge {\omega^c}_b.
  \label{Rabdef}
\end{equation}
We will refer to this simply as the curvature (of the spin connection), and its components can be related to the conventional Riemann curvature components via
\begin{equation}
  {R^\alpha}_{\rho\mu\nu}={{R^a}_b}_{\mu\nu} e^\alpha_a e^b_\rho.
  \label{curv2}
\end{equation}
Armed with these ingredients, we may write the action for Einstein-Cartan gravity in $d$ spacetime dimensions, which is given by
\begin{equation}
  S_{\rm EC}[e,\omega]=\frac{1}{32\pi G_N}
  \int_{M} \epsilon_{a_1a_2a_3\ldots a_{d-2}bc}\,e^{a_1} \wedge e^{a_2}
  \wedge\ldots \wedge e^{a_{d-2}}\wedge\left(
  R^{bc}(\omega)-\lambda e^c\wedge e^d\right),
  \label{SEC}
\end{equation}
where $G_N$ is Newton's constant, and $\lambda$ is related to the cosmological constant. We obtain two field equations from this action, given that we may vary with respect to the vielbein or spin connection. Restricting to four spacetime dimensions, for example, varying with respect to ${\omega^a}_b$ yields the zero torsion condition of eq.~(\ref{EC1}). Instead varying with respect to $e^a$ yields
\begin{equation}
  \epsilon_{abcd}e^b \wedge R^{cd}=2\lambda\epsilon_{abcd}e^b\wedge e^c\wedge e^d,
  \label{EC2}
\end{equation}
which is equivalent to the vacuum Einstein equation. 

\subsection{Gravity in 2+1 dimensions}
\label{sec:3d}

In 2+1 dimensions, the Einstein-Cartan action of eq.~(\ref{SEC}) implies the equation
\begin{equation}
R^{ab}=3\lambda e^a\wedge e^b.
\label{EOM2+1}
\end{equation}	
Thus, the curvature of an Einstein metric is constant, such that there are no propagating degrees of freedom in three-dimensional gravity\footnote{An extreme case of eq.~(\ref{EOM2+1}) is that of vanishing cosmological constant, for which one finds that spacetime must be locally flat.}. Nevertheless, solutions can be globally non-trivial (see e.g. ref.~\cite{Garcia-Diaz:2017cpv} for a useful compendium), and three-dimensional massless or massive gravity has previously proven useful as a testing ground for ideas relating to the double copy.~\cite{CarrilloGonzalez:2019gof,Gonzalez:2022otg,Gonzalez:2021bes,CarrilloGonzalez:2022ggn,Emond:2022uaf,Moynihan:2021rwh,Moynihan:2020ejh,Burger:2021wss,Alkac:2021seh,Alkac:2022tvc}.
Previous studies have focused on the Einstein--Hilbert formulation, and we here wish to provide new insights based on alternative formulations, in which gravity looks much more similar to a suitable non-abelian gauge theory. Indeed, there is a well-known formulation of gravity in 2+1 dimensions as a Chern--Simons theory, where the gauge symmetry corresponds to the Poincar\'{e} group ISO(2,1). This is known as {\it Chern--Simons--Witten (CSW) theory}, and the relevant gauge field is then valued in the Lie algebra of the Poincar\'{e} group~\cite{Witten:1988hc,Witten:1989sx}. The latter has conventional momentum and Lorentz generators $\{{\bm{P}}_a\}$ and $\{{\bm{J}}_{ab}\}$ respectively, satisfying the commutation relations
\begin{align}
\left[\bm{P}_a,\bm{P}_b\right]&=0,\notag\\
 \left[\bm{J}_{ab},\bm{P}_c\right]&=\eta_{ac}\bm{P}_b-\eta_{bc}\bm{P}_a\notag\\
\quad\quad\left[\bm{J}_{ab},\bm{J}_{cd}\right]&=\eta_{ac}\bm J_{bd}+\eta_{bd}\bm J_{ac}-\eta_{ad}\bm J_{bc}-\eta_{bc}\bm J_{ad},
\label{Palgebra}
\end{align}
and where ${\bm{J}}_{ab}=-{\bm{J}}_{ba}$. As shown in ref.~\cite{Witten:1988hc}, it is possible to equip this Lie algebra with an invariant bilinear form $\langle\cdot,\cdot\rangle$, whose action on the generators is
\begin{equation}
    \left<\bm{P}_a,\bm{P}_b\right>=0,\quad \left<\bm{J}_{ab},\bm{P}_c\right>=\epsilon_{abc},\quad \left<\bm J_{ab},\bm J_{cd}\right>=0.
\label{bilinear}
\end{equation}
Denoting an abstract generator ${\bm P}_a$ or ${\bm J}_{ab}$ by ${\bf T}_I$, the gauge field then has generic form 
\begin{equation}
{\bm{A}}=A^I{\bf T}_I = e^a{\bm{P}}_a+\frac12\omega^{ab}{\bm{J}}_{ab},
\label{Adef}
\end{equation}	
where we have included a conventional factor of $1/2$ in the second term on the right-hand side. The action for the theory can then be written as
\begin{equation}
    S_\text{CSW}=\frac{1}{2}\int_M \left<\bm{A}\wedge \d\bm{A}+\frac{2}{3}\bm{A}\wedge\bm{A}\wedge\bm{A}\right>.
    \label{SCSW}
\end{equation}
This may be shown to be equivalent to the three-dimensional case of the Einstein-Cartan action of eq.~(\ref{SEC}), upon substituting eq.~(\ref{Adef}) and using eqs.~(\ref{Palgebra}, \ref{bilinear}). We see that the CSW gauge field unifies the two fields (vielbein and spin connection) of the Einstein-Cartan theory into a single object.

Equation~(\ref{SCSW}) may be compared with the usual action for non-abelian Chern--Simons theory, with a compact gauge group $G$:
\begin{equation}
S_\text{CS}=\int_M {\rm Tr}\left({\bm{A}}\wedge \d{\bm{A}}+\frac23{\bm{A}}\wedge{\bm{A}}\wedge{\bm{A}}
\right).
\label{SCS} 
\end{equation}	
This is identical in form (up to irrelevant normalisation factors) with eq.~(\ref{SCSW}), where the relationship between the theories amounts to reinterpreting adjoint indices in the gravity theory in terms of colour indices, and replacing the invariant bilinear form of ISO(2,1) with the Killing form of the non-abelian gauge group. Thus, in the CSW formulation of gravity, there can be no ``double copy" relationship with the corresponding gauge theory. Rather, the gravity theory becomes a precise doppelg\"{a}nger of the gauge theory, allowing us to simply reinterpret indices in proceeding from one theory to another. We can then ask how the conventional double copy emerges out of this correspondence, and we explore this in more detail in the following section.

Before doing so, let us note that the vacuum field equations of Chern--Simons theory demand that the field strength
\begin{equation}
{\bm F}=\d{\bm{A}}+{\bm{A}}\wedge{\bm{A}}\
\label{Fdef}
\end{equation}
associated with the gauge field ${\bm{A}}$ vanishes. In the gravity case, an explicit calculation yields
\begin{align}
{\bm F}&=\left(\d e^a+{\omega^a}_b\wedge e^b\right)
{\bm{P}}_a+\frac12\left(\d\omega^{ab}+{\omega^a}_c\wedge\omega^{cb}\right){\bm{J}}_{ab}\notag\\
&=T^a[e,\omega]{\bm{P}}_a+\frac12 R^{ab}[\omega]{\bm{J}}_{ab}.
\label{FCSW}
\end{align}
Here we have recognised the torsion and curvature of eqs.~(\ref{Tadef}, \ref{Rabdef}), and the vanishing of the field strength then implies the zero torsion and curvature conditions found before.
Under a gauged Lorentz transformation $\Lambda$, the components of the gauge field transform as
\begin{equation}
    \begin{aligned}
        e^a \longrightarrow {\Lambda^a}_b e^b,\hspace{2em} {\omega^a}_b \longrightarrow {\Lambda^a}_c{\omega^c}_d {(\Lambda^{-1})^d}_b+{\Lambda^a}_c {(\d \Lambda^{-1})^c}_b.
    \end{aligned}
\end{equation}
For future reference, we also note that under a gauged translation generated by the algebra element $\rho^a\bm{P}_a$, the transformation will be
\begin{equation}
    e^a\longrightarrow e^a-\d\rho^a -{\omega^a}_b \rho^b,\hspace{2em} {\omega^a}_b\longrightarrow {\omega^a}_b.
    \label{eatrans}
\end{equation}
Note that this transformation in general means that the transformed $e^a$ is not necessarily a physical (invertible) vielbein field. One can thus think of this transformation as mixing the physical information about the spacetime between the degrees of freedom in $e^a$ and $\rho^a$. 

%%%%%%%%%%%%%%%%%%%%

\section{Gravity as a doppelg\"{a}nger in 2+1 dimensions}
\label{sec:2+1}

In the previous section, we have reviewed the CSW formulation of gravity, in which the gravitational action can itself be cast in the form of a gauge theory. In this section, we wish to interpret how this gives rise to the conventional three-dimensional double copy between gauge theory and gravity. For this to work, however, we must consider the coupling of gravity to matter. We note that it is not known in general how to couple arbitrary matter sources to gauge theory and gravity and preserve the double copy (see e.g. refs.~\cite{Ridgway:2015fdl,Carrillo-Gonzalez:2017iyj,Easson:2021asd,Easson:2022zoh} for important works). Likewise, equivalence between different formulations of gravity can itself be disrupted when trying to include matter. Here, we will restrict ourselves to sources which are highly localised, such that we may consider gravity coupled to discrete worldlines. Our aim is to cast gauge theory and gravity couplings in a form that looks almost identical, and to see how this gives rise to known double copy properties. 

\subsection{Wilson lines and loops}
\label{sec:Wilson}

Given a gauge field ${\bm{A}}_\mu$, there is a natural way to couple it to worldlines, namely through the {\it Wilson line operator}
\begin{equation}
\Phi_{\cal C}[{\bm{A}}]={\cal P}\exp\left[g\int_{\cal C} dx^\mu
{\bm{A}}_\mu\right]={\cal P}\exp\left[g {\bf T}^a\int_{\cal C} d\tau \dot{x}^\mu
A^a_\mu\right],
\label{Phidef}
\end{equation}	
where $g$ is the coupling constant, $\tau$ the proper time along a given contour ${\cal C}$, and the dot denotes differentiation with respect to $\tau$. The exponent is matrix-valued in the vector space corresponding to a particular representation of the gauge group, which is fixed in eq.~(\ref{Phidef}) by the representation of the generators $\{{\bf T}^a\}$. The exponential itself is defined by its Taylor expansion, and the ${\cal P}$ symbol in eq.~(\ref{Phidef}) denotes that the generators should then be {\it path-ordered} i.e. ordered according to their parameter distance along ${\cal C}$. In gauge theories, Wilson lines are associated with the transportation of gauge information between different points in the underlying manifold, and are thus a crucial ingredient for constructing gauge-covariant or gauge-invariant observables. They also arise as dressing factors for scattering amplitudes in certain kinematic limits, such as when fast-moving particles emit low energy (``soft") radiation. In gravity, the relevant operator describing soft graviton emission is given by~\cite{Naculich:2011ry,White:2011yy} 
\begin{equation}
\Phi^{\rm grav.}_{\cal C}=\exp\left[\frac{\kappa}{2}
m\int_{\cal C} dx^\mu \dot{x}^\nu h_{\mu\nu},
\right],
\label{Phigrav}
\end{equation}
where $h_{\mu\nu}$ is the graviton field defined by 
\begin{equation}
g_{\mu\nu}=\eta_{\mu\nu}+\kappa h_{\mu\nu},
\label{hdef}
\end{equation}
with $\kappa=\sqrt{32\pi G_N}$ the gravitational coupling in terms of Newton's constant $G_N$, and $m$ the mass of the particle. For a straight-line contour such that
\begin{equation}
x^\mu=u^\mu \tau,
\label{xparam}
\end{equation}	
where $\beta^\mu$ is the 4-velocity, one finds that the operator of eq.~(\ref{Phigrav}) is obtained from eq.~(\ref{Phidef}) by: (i) replacing the field $A_\mu^a$ with $h_{\mu\nu}$ (itself involving replacing a colour index with a spacetime one); (ii) replacing coupling constants according to $g\rightarrow \kappa/2$; (iii) replacing the colour generator ${\bf T}^a$ with the momentum $p^\mu=m\beta^\mu$. As argued in refs.~\cite{Melville:2013qca,Luna:2016idw,Alfonsi:2020lub}, these replacements --  which also generalise to the case of massless particles -- correspond to the known double copy for scattering amplitudes (see also ref.~\cite{Oxburgh:2012zr} for a non-Wilson line treatment of soft radiation and the double copy). In these previous studies, eq.~(\ref{Phidef}) has only been considered in the context of non-abelian gauge theory with a compact gauge group. Here, however, we have reviewed that gravity can itself be written as a gauge theory in 2+1 dimensions. Thus, we must be able to take ${\bm{A}}_\mu$ in eq.~(\ref{Phidef}) as the Chern--Simons--Witten gravity field. In that language, the form of the coupling to a worldline looks identical to that in non-abelian gauge theory, such that there is a manifest doppelg\"{a}nger relationship between the two theories. However, one must then show that the CSW coupling reproduces the known Wilson line operator double copy implicit in eq.~(\ref{Phigrav}). 

In fact, there are numerous ways to show that the CSW Wilson line is consistent with eq.~(\ref{Phigrav}). For concreteness, we will focus on the explicit case of a gauge-invariant {\it Wilson loop}, in which the contour ${\cal C}$ is taken to be a closed curve. Given that both non-abelian Chern--Simons theory and three-dimensional gravity are topological field theories, Wilson loops and their expectation values have been widely studied due to their ability to encode topological invariants (see e.g. ref.~\cite{Baez:1995sj} for a comprehensive review).  Physically, Wilson loops describe the phase experienced by a particle traversing a closed curve, and to be scalar-valued and gauge-invariant must then include a trace in the given representation ${\cal R}$ of the gauge group:
\begin{equation}
W_{\cal C}[{\bm{A}}]={\rm Tr}_{\cal R}{\cal P}\exp\left[
i\oint_{\cal C} dx^\mu {\bm{A}}_{\mu}\right].
\label{Wdef}
\end{equation}	
In order to make contact with the gravitational Wilson loop, the relevant representation of the Poincar{\'e} group will be that of a massive spinless particle. This will necessarily be infinite dimensional which complicates the direct implementation of the trace, so we follow an alternative approach suggested in \cite{Witten:1989sx} and employed in \cite{Castro:2020smu,Ammon:2013hba} within the context of AdS CSW gravity. The basic idea is to consider the test particle on the wordline as a quantum mechanical system whose Hilbert space provides a representation space for the gauge group. That is, we may introduce coordinates $\{q^a\}$ and their conjugate momenta $\{p^a\}$ associated with an internal space at each point on the worldline, and such that a local Poincar\'{e} transformation at each point corresponds to
\begin{equation}
q^a\rightarrow {\Lambda^a}_b q^b+\rho^a,\quad p^a\rightarrow {\Lambda^a}_b q^b,
\label{Poincare}
\end{equation}
where ${\Lambda^a}_b$ is a Lorentz transformation matrix. We can couple a background CSW gauge field to the internal Poincar\'{e} degrees of freedom by introducing the worldline covariant derivative
\begin{equation}
D q^a = \d q^a +{\omega^a}_b q^b + e^a,
\label{Dqa}
\end{equation}	
where ${\omega^a}_b$ and $e^a$ are the background spin connection and vielbein. In coordinate notation this reads
\begin{equation}
D_\mu q^a = \partial_\mu q^a +{(\omega_\mu)^a}_b q^b + e_\mu^a,
\label{Dqa2}
\end{equation}	
and the locally gauge invariant action coupling the test particle to the CSW field is given by
\begin{align}
S[q^a,p_a;{\bm{A}}]&=\int_{\cal C} d\tau \left[p_a \dot{x}^\mu(\partial_\mu q^a+{(\omega_\mu)^a}_b q^b + e_\mu^a)-\xi(p_a p^a+m^2)\right],
\label{Sdef}
\end{align}	
where the second term, involving a Lagrange multiplier $\xi$, implements the mass-shell constraint. If we have an open Wilson line with contour ${\cal C}$, there will be an initial state of the $|i\rangle$ of the system $\{q^a,p_a,\xi\}$ at one end of the contour, and a final state $|f\rangle$ at the other. The expectation value of the Wilson line sandwiched between these states can then be interpreted as a transition amplitude from $|i\rangle$ to $|f\rangle$ in the presence of the background CSW gauge field, which has the path-integral representation
\begin{equation}
\left\langle f\left| {\cal P}\exp\left(i\int_{\cal C} dx^\mu {\bm{A}}_\mu\right)\right| i\right\rangle=
\int {\cal D} q^a{\cal D} p_a{\cal D}\xi \,e^{iS[q^a,p_a;{\bm{A}}]}.
\end{equation}	
On the right-hand side, one integrates over all possible values of the dynamical degrees of freedom at all points on the worldline, subject to the appropriate boundary conditions at the end-points corresponding to $|i\rangle$ and $|f\rangle$. For a closed loop ${\cal C}$, one may integrate over all values of the field at {\it all} points in the loop. If one fixes a point to be the common start and end of the loop, this amounts to summing over all states at that point, and thus to carrying out the trace in eq.~(\ref{Wdef}). That is, one has~\cite{Witten:1989sx,Castro:2020smu,Ammon:2013hba}
\begin{equation}
{\rm Tr}_{\cal R}{\cal P}\exp\left[
i\oint_{\cal C} dx^\mu {\bm{A}}_{\mu}\right]=\int {\cal D}q^a {\cal D}p_a{\cal D}\xi\,
\exp\left[
i\oint_C d\tau p_a\dot{x}^\mu(\partial_\mu q^a+{(\omega_\mu)^a}_b q^b+e_\mu^a)-\xi(p_ap^a+m^2)\right],
\label{Wint}
\end{equation}
and the usefulness of this expression is that one may explicitly carry out the path integrals appearing on the right-hand side. Given that the Wilson line corresponds to the phase experienced by a particle following a classical trajectory, we may carry out the integrals using the saddle point approximation, commencing with the integral over $q^a$. Varying the action with respect to the latter gives the the equation of motion
\begin{equation}
    \dot{x}^\mu\left[\partial_\mu p^a+{(\omega_\mu)^a}_bp^b\right]=0,
\end{equation}
which when substituted back into the path integral yields
\begin{equation}
    W[C]=\int[\mathcal{D}p\mathcal{D}\xi] \exp\left[i\oint_C d\tau \Big( \dot{x}^\mu p_a e_\mu^a-\xi\left(p_ap^a+m^2\right)\Big) \right].
\end{equation}
Next, the equation of motion for $p^a$ is found to be
\begin{equation}
    p^a=\frac{1}{2\xi}e_\mu^a \dot{x}^\mu,
\end{equation}
such that we are left with
\begin{equation}
    W[C]=\int[\mathcal{D}\xi] \exp\left[i\oint_Cd\tau\Big(\frac{1}{4\xi}\dot{x}^\mu \dot{x}^\nu e_\mu^ae_\nu^b\eta_{ab}-\xi m^2 \Big)\right].
    \label{Wform2}
\end{equation}
The field $e_\mu^a$ on the worldline is the pullback of the vielbein field associated with the background bulk CSW gauge field. From eq.~(\ref{metricdef2}), we may thus recognise the conventional metric tensor, such that eq.~(\ref{Wform2}) becomes
\begin{equation}
    W[C]=\int[\mathcal{D}\xi] \exp\left[i\oint_C d\tau\left(\frac{1}{4\xi}g_{\mu\nu}\dot{x}^\mu \dot{x}^\nu-\xi m^2\right) \right].
    \label{Wform3}
\end{equation}
At this stage, one may gauge-fix $\xi$ and substitute the graviton definition of eq.~(\ref{hdef}) to yield the Wilson line operator of eq.~(\ref{Phigrav}), where the advantage of this approach is that the operator works manifestly for either massless or massive test particles. Alternatively, for non-zero mass one may eliminate $\xi$ using its equation of motion to get 
\begin{equation}
    W[C]=\text{exp}\left[-im\oint_C d\tau\sqrt{-g_{\mu\nu}\dot x^\mu\dot x^\nu}\right].
\label{Wform4}
\end{equation}
This makes clear that the Wilson line operator is associated with the path-length of the worldline. Again, substituting eq.~(\ref{hdef}) and expanding in $\kappa$ yields the known Wilson line operator of eq.~(\ref{Phigrav}), as noted previously in e.g. ref.~\cite{Alfonsi:2020lub}.

Above, we have argued that the known double copy of Wilson line operators for soft gluon / graviton radiation emerges from casting both theories in the same language -- that of a non-abelian Chern--Simons gauge theory. It is possible, however, to make the common structure of the coupling of gauge fields to matter much more explicit in the two theories, as we will now see.

\subsection{Point particle couplings in gauge theory and gravity}
\label{sec:particle}

As reviewed in e.g. ref.~\cite{Goldberger:2016iau}, the current for a pointlike colour charge in a non-abelian gauge theory can be written (in $D$ spacetime dimensions) as
\begin{equation}
j^{\mu\,a}(y^\mu)=\int d\tau c^a(\tau) \dot{x}^\mu \delta^D(y^\mu-x^\mu(\tau)),
\label{jmua}
\end{equation}
where the delta function localises onto a particular worldline ($x^\mu(\tau)$), and $c^a(\tau)$ is a colour vector associated with each point along the worldline. Covariant current conservation $D_\mu j^{\mu\,a}$ then implies 
\begin{equation}
\frac{d c^a}{d\tau}=g f^{abc} \dot{x}^\mu(\tau) A_\mu^b(\tau) c^c(\tau),
\label{Wong1}
\end{equation} 
where $\{f^{abc}\}$ are the structure constants of the gauge group. The particle trajectory itself obeys the non-abelian generalisation of the Lorentz force law:
\begin{equation}
m\frac{d^2 x^\mu}{d\tau^2}=g c^a {F^{a\,\mu}}_\nu \dot{x}^\nu.
\label{Wong2}
\end{equation}	
Collectively, eqs.~(\ref{Wong1}, \ref{Wong2}) are known as the {\it Wong equations}, having first been derived in ref.~\cite{Wong:1970fu}. The question then arises of how to derive these equations from a suitable point particle action, and various prescriptions have appeared in the literature~\cite{Balachandran:1976ya,Balachandran:1977ub}. 

The approach of ref.~\cite{Balachandran:1977ub} uses as its dynamical variable a faithful representation of the gauge group attached to the worldline, such that a group element associated with proper time $\tau$ can be written as $\bm{g}(\tau)$. For a given gauge group $G$, the quantity
\begin{equation}
    \bm{g}^{-1}\bm{D}_\mu \bm{g}=
    \bm{g}^{-1}\Big(\partial_\mu +A_\mu^b\bm{T}_b
    \Big)\bm{g}
    \label{GMC}
\end{equation}
is known as the {\it gauged Maurer-Cartan form}, and is a one-form valued in the Lie algebra of $G$. An action for Yang-Mills theory coupled to the worldline can then be obtained by taking the inner product of this form -- defined at all points on the worldline -- with a Lie algebra element 
\begin{equation}
    \bm{K}=K^a \bm{T}_a,
    \label{Kdef}
\end{equation}
where the form of the action itself is
\begin{equation}
    S=\int ds +\int d\tau \dot x^\mu\left< K^a \bm{T}_a,\bm{g}^{-1}(\tau) \left( \partial_\mu + A^b_\mu[x(\tau)] \bm{T}_b\right)\bm{g}(\tau) \right>.
    \label{SYM}
\end{equation}
One may think of the role of $\bm{K}$ as fixing the particular representation of $G$ that acts on the worldline (in other words, the ``type" of colour charge that is being coupled to Yang-Mills theory). To make this precise, one may identify 
\begin{equation}
c^a\bm{T}_a=\bm{g}\bm{K}\bm{g}^{-1},
\label{cadef}
\end{equation}
and then obtain equations of motion by varying the action of eq.~(\ref{SYM}) with respect to both $\bm{g}$ and the worldline coordinate $x^\mu(\tau)$. As shown in ref.~\cite{Balachandran:1977ub}, this yields the Wong equatons of eqs.~(\ref{Wong1}, \ref{Wong2}). Thus, eq.~(\ref{cadef}) tells us that the Lie algebra element $\bm{K}$ in eq.~(\ref{Kdef}) determines the colour vector carried by Wong's moving charged particle. 

Given that the above arguments are independent of the gauge group, they may also be applied directly to CSW gravity. In that case, one inserts the Poincar{\'e} valued gauge field into the coupling action, such that ${\bm g}\equiv (\Lambda,q)$ is an element of the Poincar{\'e} group (with Lorentz and translation parameters $\Lambda$ and $q$ respectively), and $\bm K$ a Poincar{\'e} algebra element: 
\begin{displaymath}
    \bm{K}=p^a\bm{J}_a+s^a\bm{P}_a,\quad \bm{J}_a=\frac12{\epsilon_{a}}^{bc}\bm{J}_{bc}.
\end{displaymath} 
In order to concretely compute the action, we use a convenient $4\times 4$ matrix representation of the group elements as in e.g. ref.~\cite{deSousaGerbert:1990yp}
\begin{equation}
    \bm{g}=\left(\begin{matrix}
        {\Lambda^i}_j & q^i\\
        0&1
    \end{matrix}\right) \hspace{3em} \bm{K}=\left( \begin{matrix}
        -p^a{{\epsilon_a}^i}_j & s^a{\delta_a}^i\\
        0&0
    \end{matrix} \right),
\end{equation}
where the generators are in turn given by
\begin{equation}
    \bm{J}_a = \left(\begin{matrix}
        (-{{\epsilon_a)}^i}_j & 0\\
        0&0
    \end{matrix}\right)\hspace{2em}\bm{P}_a = \left(\begin{matrix}
        0 & (\delta_a)^i\\
        0&0
    \end{matrix}\right).
\end{equation}
The second term in eq.~(\ref{SYM}) represents the coupling of the gauge field to the worldline, and in the gravity case this now reads
\begin{equation}\label{wlaction}
    \begin{aligned}
        S_\text{coupling} &= \int_\gamma \left<p^a\bm{J}_a + s^a\bm{P}_a, \left(\begin{matrix}
            \Lambda^{-1} & -\Lambda^{-1}q\\
            0&1
        \end{matrix}\right)\cdot\left[\left(\begin{matrix}
            d\Lambda & dq\\
            0&0
        \end{matrix}\right)+\left(\begin{matrix}
            \omega^a\epsilon_a & e^a\delta_a\\
            0&0
        \end{matrix}\right)\cdot\left(\begin{matrix}
            \Lambda & q\\
            0&1
        \end{matrix}\right)\right]\right>\\
        &=\int_\gamma \left< p^a\bm{J}_a+s^a\bm{P}_a, \frac{1}{2}{(\Lambda^{-1})^i}_k \left[d{\Lambda^k}_j+{\omega^k}_l{\Lambda^l}_j\right]{(-\epsilon^a)^j}_i \,\bm{J}_a +{(\Lambda^{-1})^a}_b Dq^b \bm{P}_a\right>.
    \end{aligned}
\end{equation}
Applying the invariant bilinear form of \eqref{bilinear}, the action becomes
\begin{equation}
    S_\text{coupling}=\int_\gamma p_a {(\Lambda^{-1})^a}_b
    \left(dq^b+{\omega^b}_cq^c+e^c\right)+\frac{1}{2}s^a\text{Tr}\left[\Lambda^{-1}\left(d\Lambda + \omega \Lambda\right)\bm{J}_a\right].
\end{equation}
In a spinless massive representation, the second term of this action associated with $s^a$ vanishes and $\Lambda$ becomes non-dynamical, allowing us to absorb it into $p_a$ and giving the action
\begin{equation}
    S_\text{coupling}^\text{massive, spinless} = \int\text{d}\tau\, p_a\left(\partial_\mu q^a+{(\omega_\mu)^a}_bq^b+e_\mu^a\right)\dot x^\mu
    \label{Scoupling}
\end{equation}
which, when supplied with a mass-shell constraint can be seen to reproduce \eqref{Sdef}. Combined with the results of the previous section, the second term that appears in eq.~(\ref{SYM}) thus indeed reproduces the known double copy of Wilson lines that couple gravitons or gluons to worldlines. Here, though, we have cast this relationship in a form that looks the same in both theories: the coupling term in eq.~(\ref{SYM}) is independent of the gauge group, suggesting that one may simply reinterpret generators and colour indices in proceeding from one theory to another. This is thus an explicit manifestation of the BCJ (or colour-kinematics) duality property, which posits that one must replace a colour algebra by a kinematic algebra upon performing the double copy. Here, this acts as a literal replacement, such that colour indices are simply reinterpreted as spacetime ones.

Careful scrutiny of the above analysis shows that it is almost -- but not quite -- true that the CSW formulation places the gauge and gravity theories on a precisely equal footing. Although the second term in eq.~(\ref{SYM}) matches up between the two types of theory, the first term on the right-hand side does not. In gauge theory, this term must be inserted by hand in order to generate the dynamics of the free particle. In gravity, however, the total length of the worldline of the particle emerges from the coupling term, given that eq.~(\ref{Scoupling}) ends up generating the Wilson line operator of eq.~(\ref{Wform4}). This can be viewed as a counterpart of the fact that in the conventional Einstein--Hilbert formulation of gravity, the double copy applies to the graviton field $h_{\mu\nu}$ rather than the full metric $g_{\mu\nu}$. Furthermore, this observation does not invalidate the fact that the CSW formulation provides a concrete setting in which interaction terms in gravity are a manifest doppelg\"{a}nger of their analogues in gauge theory, with a strict replacement of colour generators by kinematic ones.

Given that the action of eq.~(\ref{SYM}) reproduces the Wong equations in gauge theory, it is natural to ask what their counterpart represents in gravity. This is the subject of the following section.

\subsection{Doppelg\"anger dynamics: Wong versus Mathisson--Papapetrou--Dixon}\label{sec:doppel_dynamics}
In the previous section, we have seen that at least in in 2+1 dimensions, interaction terms in gauge theory and gravity can be written such that there is a manifest doppelg\"anger relationship between the two theories. For colour degrees of freedom, this leads to the Wong equations ~(\ref{Wong1}, \ref{Wong2}), and it is now pertinent to ask what these indices represent when interpreted gravitationally. The Wong equations describe the motion of a coloured particle in a non-Abelian background and, as we will now show, the exact same formalism describes a \textit{spinning} particle moving on a curved background, governed by the Mathisson--Papapetrou--Dixon (MPD) equations~\cite{Mathisson:1937zz,Papapetrou:1951pa,Dixon:1970zza}. Furthermore, our arguments will apply to the general case of $D$ spacetime dimensions, rather than being limited to the 2+1 case. To see this explicitly, it is convenient to rewrite the colour charge and field strength in the adjoint (matrix) representation,
\begin{equation}
Q^{ab} = f^{abc} c^c,
\qquad
F_{\mu\nu}^{ab} = f^{abc} F_{\mu\nu}^c,
\label{Qadj}
\end{equation}
so that the commutator structure of the non--abelian field is explicit.
With these definitions eqs.~(\ref{Wong1},~\ref{Wong2}) become
\begin{equation}\label{WongAdj}
\frac{dQ^{ab}}{d\tau}
= \dot{x}^\mu\bigl[A_\mu, Q\bigr]^{ab} = \dot{x}^\mu A_\mu^{ac}Q^{cb} - \dot{x}^\mu A_\mu^{bc}Q^{ca}
\end{equation}
and
\begin{equation}
    m\frac{d^2x^\mu}{d\tau^2} = g Q^{ab} F^{ab\mu}{}_{\nu} \,
    \dot x^\nu.
\end{equation}
Notice that the right-hand side of the first equation involves a single contraction of two adjoint
indices, leaving a free pair $(ab)$ that mirrors the antisymmetric index pair
$(\rho\sigma)$ of a local Lorentz bivector. We are now free to interpret the indices $(ab)$ as belonging to either $SU(N)$ or $SO(D-1,1)$, describing a gauge or gravitational theory, respectively. However, it is notationally convenient to make the explicit identifications 
\begin{equation}
Q^{ab} \longrightarrow S^{ab}+L^{ab},
\qquad
g F_{\mu\nu}^{ab} \longrightarrow
\frac{1}{2} R^{ab}_{\mu\nu},
\qquad
A_\mu^{ab} \longrightarrow \frac12\omega_\mu^{ab},
\label{DCmap}
\end{equation}
where $S^{ab}$ is the intrinsic spin-tensor, $L^{ab}$ the orbital tensor and $\omega_\mu^{ab}$ the spin-connection. These replacements are such that we land on the conventional form of the MPD equations 
\begin{equation}\label{MPDeqs}
\frac{Dp^\mu}{d\tau} = -\frac{1}{2} R^\mu{}_{\nu ab} S^{ab} \dot x^\nu,~~~~~~
\frac{dS^{ab}}{d\tau} =  \dot{x}^\mu(\omega_\mu^{ac}S_c^{~b} - \omega_\mu^{bc}S_c^{~a}) + p^{a}u^{b}-p^{b}u^{a},
\end{equation}
and essentially amount to simply replacing the covariant derivative, but in its slightly unusual adjoint form: 
\begin{equation}
D_\mu = \pd_\mu + gA_\mu^{ab}{\bf T}^{ab},~~~~~[D_\mu,D_\nu] = gF_{\mu\nu}^{ab}{\bf T}^{ab}.
\label{Dmuadj}
\end{equation}
Here we have defined the dual colour generators
\begin{equation}
{\bf T}^{ab}=f^{abc}{\bf T}^c.
\label{dualcol}
\end{equation}
Upon making the above replacements together with the additional replacement
\begin{displaymath}
{\bf T}^{ab}\rightarrow M^{ab},
\end{displaymath}
where $M^{ab}$ is a generator of an internal Lorentz algebra,
eq.~(\ref{Dmuadj}) simply becomes the gravitational covariant derivative, now identifying the colour indices with tangent bundle indices:
\[
D_\mu = \pd_\mu + \frac12\omega_\mu^{ab}M^{ab},~~~~~[D_\mu,D_\nu] = \frac12R_{\mu\nu}^{ab}M^{ab}.
\]
Equation~(\ref{DCmap}) is the dynamical analogue of the colour-to-kinematics replacement rules that produced the operator-level double copy in section~\ref{sec:Wilson}.

Our remarks above have been for the $D$ spacetime dimensions. Returning to the explicit example of 2+1 dimensions, both the Wong and MPD equations can be derived directly from the CSW coupling by extending the action of eq.~(\ref{Sdef}) to include a bivector $Q^{\bar{a}\bar{b}}$, and we use barred indices to make the distinction clear: if we unbar $\bar{a}$, it becomes a local Lorentz (tangent bundle) index, otherwise it should be thought of as a gauge-group colour index. The gauged Maurer--Cartan action of eq.~(\ref{Sdef}) then becomes
\begin{equation}
S\bigl[q,p,\Sigma;{\bm{A}}\bigr]
:=
\int_{\cal C} d\tau 
\Bigl[
p_a D_\tau q^a
+ \frac{1}{2} Q^{\bar{a}\bar{b}}\Omega_{\bar{a}\bar{b}}
- \xi(p^2+m^2),
\Bigr],
\label{SpinAction}
\end{equation}
which is a generalisation of the action of eq. \eqref{wlaction}, where 
\begin{equation}
\Omega_{\bar{a}\bar{b}} = \bigl(g^{-1}D_\tau g\bigr)_{\bar{a}\bar{b}}. 
\end{equation}
in gauge theory. In gravity, one replaces the group element $g$ with an element of the Lorentz group. The resulting equations of motion from eq.~(\ref{SpinAction}) are
\[
D_\tau Q^{\bar{a}\bar{b}} = 0,~~~~~D_\tau p^a= Q^{\bar{a}\bar{b}}F_{\bar{a}\bar{b}}^{~~ab}u_b,~~~~~D_\tau q^a = 2\xi p^a,
\]
%where we have used
%\[
%\frac{\partial \Omega_{ab}}{\partial x^\mu} = 
%\]
If we unbar the indices here, and identify $Q^{ab}$ as the \textit{total} angular momentum tensor, then these are the MPD equations. If not, then they are the Wong equations. From this perspective the same action yields both the Wong and MPD systems, making (\ref{SpinAction}) the parent ``master''
action whose equations of motion admit two doppelg\"{a}nger incarnations. Having seen how actions and equations of motion can be mapped between gauge and gravity theories, let us now turn to their explicit solutions.

\subsection{Point source solutions in 2+1 dimensional gravity}\label{sec:pointsource}

The solutions of the Einstein equations sourced by point masses and spins in 2+1 dimensions have been studied extensively in both metric and first-order formalisms (see e.g. refs.~\cite{Garcia-Diaz:2017cpv,deSousaGerbert:1990yp}). It is well understood that in the absence of a cosmological constant, metric solutions are automatically Riemann flat and do not contain propagating degrees of freedom. Consequently, point masses no longer source Schwarzschild-like potentials and instead introduce topological defects into the spacetime manifold in the form of conical singularities. The metric solution for a single point particle of mass $M$ can be written in polar coordinates as \cite{Garcia-Diaz:2017cpv}
\begin{equation}
    \d s^2=-\left( \d t+4 G_N S\,\d\phi \right)^2 +\frac{\d r^2}{\left(1-4 G_N M\right)^2}+r^2 \d\phi^2 \hspace{2em} t\in[-\infty,\infty],r\in[0,\infty],\phi\in[0,2\pi).
\end{equation}
By rescaling the radial and azimuthal coordinates, this metric can be brought to a form which is explicitly Minkowski,
\begin{equation}
    t\rightarrow \tilde t=t+4 G_N S\,\phi, \hspace{2em} r\rightarrow \tilde r=\frac{r}{1-4 G_N M}, \hspace{2em}\phi\rightarrow \tilde \phi =\left(1-4 G_N M\right)\phi,
\end{equation}
but this transformation has a non-trivial topological effect, altering the boundary conditions on the azimuthal coordinate $\tilde\phi\in \left[0, 2\pi(1-4G_N M)\right)$. This shows the presence of a conical singularity in the spacetime which has no effect on the curvature but does induce non-trivial phase factors for semi-classical test particles looping around the source, analogous to the Aharonov--Bohm effect in QED \cite{Ortiz:1991gx}.\par

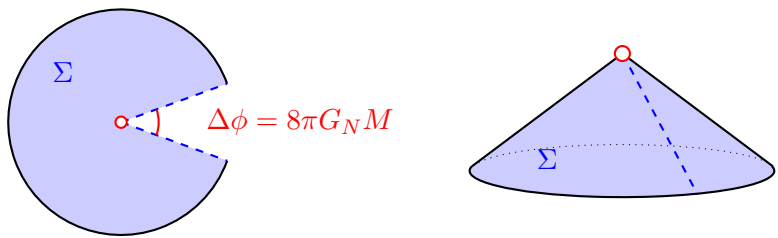
\begin{figure}[h]
    \centering
    \begin{minipage}[r]{0.45\linewidth}
        \centering
        \begin{tikzpicture}[scale=0.5]
          % Parameters
            \def\r{3}         % Radius
            \def\a{20}        % Start angle
            \def\b{340}       % End angle
          % Fill the circular segment
            \fill[blue!20] (0,0) ++(\a:\r) arc (\a:\b:\r) -- (0,0) -- cycle;
          % Outline
            \draw[thick] (0,0) ++(\a:\r) arc (\a:\b:\r);  % Arc
            \draw[thick,blue,dashed] (0,0) -- ++(\a:\r);  % Chord start
            \draw[thick,blue,dashed] (0,0) -- ++(\b:\r);  % Chord end
            \draw[thick,red] (0,0) ++(\a:\r/3) arc (\a:\b-360:\r/3);
            \fill[white] (0,0) circle (\r /20);
            \draw[thick,red] (0,0) circle (\r /20);

            \node[red] at (0:2) [right] {\(\Delta \phi = 8 \pi G_N M \)};
            \node[blue] at (-1,0.8) [above left] {\(\Sigma\)};
        \end{tikzpicture}
    \end{minipage}\hspace{-5em}
    \begin{minipage}[l]{0.45\linewidth}
        \centering
        \begin{tikzpicture}[scale=0.5]
            \def\h{3}     % height
            \def\rx{4}    % x-radius of base ellipse
            \def\ry{0.7}  % y-radius of base ellipse
            \def\off{1.5}

            \path[fill=blue!20] 
                (-\rx,0+\off) -- (0,\h+\off) -- (\rx,\off)
                arc[start angle=10, end angle=-190, x radius=\rx+0.05, y radius=\ry] -- cycle;

            \draw[dotted] (\rx,\off+0) arc[start angle=10, end angle=170, x radius=\rx+0.05, y radius=\ry];

            \draw[thick] (-\rx,\off+0) -- (0,\off+\h);
            \draw[thick] (\rx,\off+0) -- (0,\off+\h);
            \draw[dashed,thick,blue] (0,\off+\h) -- (2,\off-0.75);

            \draw[thick] (-\rx,\off+0) arc[start angle=170, end angle=370, x radius=\rx+0.05, y radius=\ry];
            \fill[white] (0,\off+\h) circle (0.2);
            \draw[thick,red] (0,\off+\h) circle (0.2);

            \node[blue] at (-2,\off+0.2) {\(\Sigma\)};
        \end{tikzpicture}
    \end{minipage}
    \caption{The conical geometry of a spatial slice $\Sigma$ induced by a point mass at the origin. The deficit angle $\Delta \phi$ is proportional to the mass. Identified edges are indicated by dashed blue lines.}
    \label{fig:enter-label}
\end{figure}

We can translate this spacetime geometry into the CSW language by identifying an appropriate dreibein and spin connection which correspond to the metric. There is a Lorentz freedom in making this identification, and we choose the following form in order to eliminate $\mathcal{O}(1)$ terms in the spin connection:
 \begin{equation}
     \begin{aligned}
         &e^0=\d t+4 G_N S \d\phi, &\hspace{2em}{\omega^1}_2=4 G_N M\d\phi,\\
         &e^1=\frac{\cos\phi\,\d r}{1-4 G_N M}-r\sin\phi\,\d\phi,&\hspace{2em}{\omega^2}_0=0,\\
         &e^2=\frac{\sin\phi\,\d r}{1-4 G_N M}+r\cos\phi\,\d\phi,&{\omega^0}_1=0.
     \end{aligned}
 \end{equation}
 We can now make use of the gauged Poincar{\'e} symmetry to perform a transformation parameterised by $ \rho^a\bm{P}_a$, under which $e^a$ transforms as in eq.~(\ref{eatrans}). Upon choosing the globally well-defined transformation parameters
 \begin{equation}
     \rho^a = \left(\begin{matrix}
         t & \frac{r\cos \phi}{1-4 G_N M} & \frac{r\sin \phi}{1-4 G_N M}
     \end{matrix}\right)^a,
 \end{equation}
 the full CSW gauge field assumes the form
 \begin{equation}
     \bm{A} = 4 G_N\left(M\bm{J}_{12}+S\bm{P}_0\right)\d\phi=\bm{U}^{-1}\d\bm{U},
     \hspace{2em}\bm{U}=\exp\left[ 4 G_N\left(M\bm{J}_{12}+S\bm{P}_0\right)\phi \right].
     \label{ACSWform}
 \end{equation}
 This gauge field is pure gauge, but possesses a non-trivial monodromy about the origin. The gauge choice has also decoupled the colour and kinematic parts of the gauge field, yielding a globally constant colour factor $M\bm{J}_{12}+S\bm{P}_0$ multiplying a solution to the abelian Chern--Simons equations of motion.

\begin{figure}
    \centering
    \includegraphics[width=0.8\linewidth]{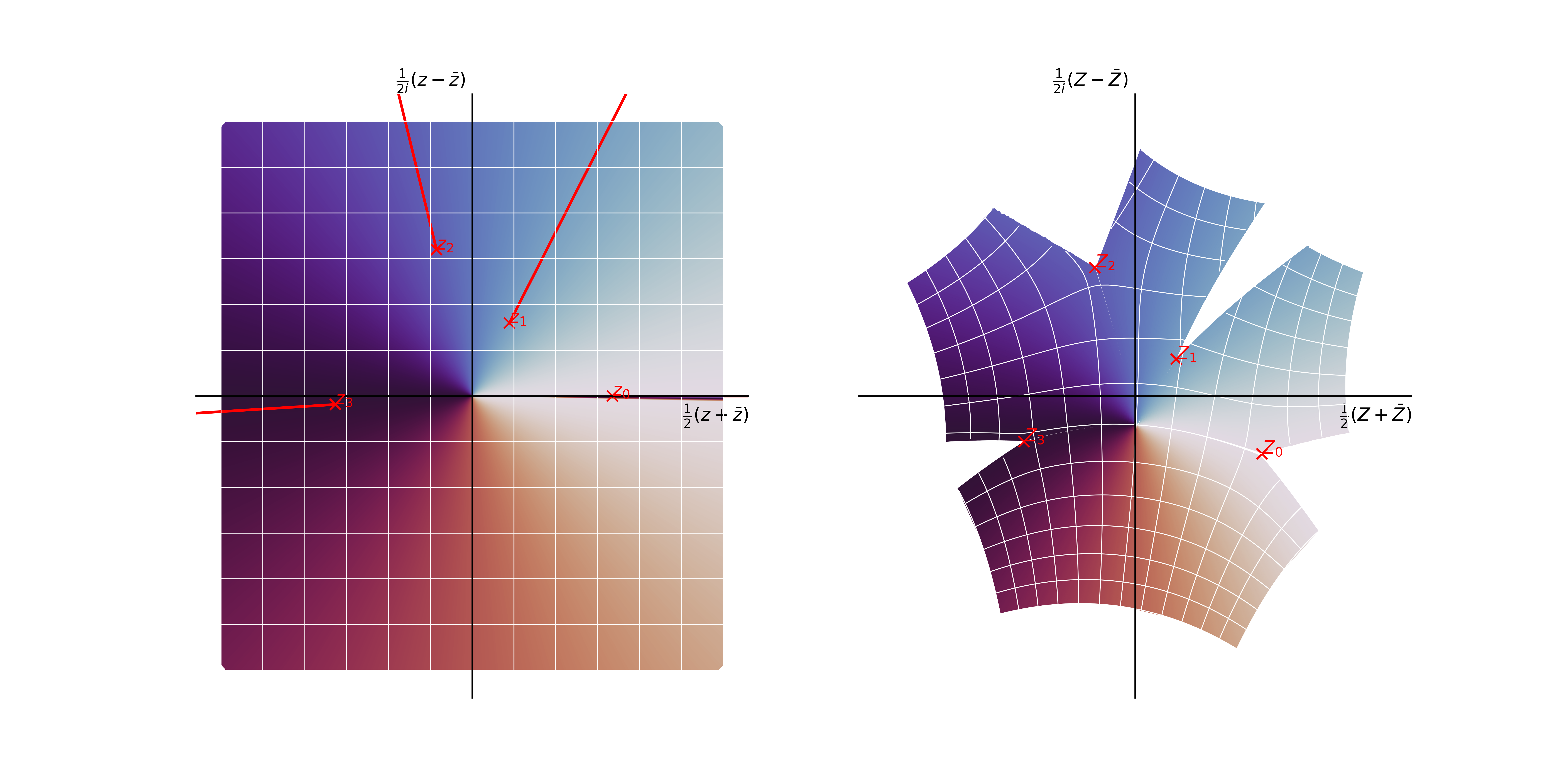}
    \caption{The change of spatial coordinates from left to right makes the metric explicitly Minkowski, but is no longer defined over the entire complex plane. Branch cuts introduce deficit angles around point sources. Identifying edges on either side of the deficit allows the surface to be wrapped into a ``multicone".}
    \label{fig:placeholder}
\end{figure}

Above, we have seen that the point mass solution of CSW gravity can be gauge-transformed so as to only involve generators in the Cartan subalgebra of the ISO(2,1) (Poincar\'{e}) group. It then becomes immediate to map this solution to non-abelian Chern--Simons theory. Given a non-abelian gauge group $G$ (e.g. SU(N)), one can simply map the generators ${\bm{J}_{12}}$ and ${\bm{P}}_0$ appearing in eq.~(\ref{ACSWform}), to generators appearing in the Cartan algebra of $G$. For SU(3), for example, one could simply map
\begin{displaymath}
{\bm{J}}_{12}\rightarrow {\bf T}^3,\quad {\bm{P}}_0\rightarrow 
{\bf T}^8
\end{displaymath}
in the conventional Gell-Mann basis. Then the resulting gauge field satisfies the usual non-abelian Chern--Simons equations, and one may reinterpret the mass and spin parameters $M$ and $S$ in terms of appropriate charges. That this is possible relies crucially on the fact that the generators appearing on the gravity side can be replicated by a corresponding subalgebra on the gauge theory side, and we return to this point in what follows. Before doing so, we note 
that the above approach can in fact be extended to a static multipole solution describing an arbitrary number of point masses and spins. To this end, we may take the multicenter solution in homogeneous coordinates as given for example in ref.~\cite{Garcia-Diaz:2017cpv}:
 \begin{equation}
     \d s^2=-\left(\d t+4 G_N\sum_n S_n \frac{\epsilon_{ij}\left(r^i-r_n^i\right)\d x^j}{|\vec r-\vec r_n|^2}\right)^2+\prod_n |\vec r-\vec r_n|^{-8  G_N M_n}\left(\d x^2+\d y^2\right)
     \label{ds2multi}
 \end{equation}
 and re-express it using complex coordinates on the $x-y$ plane:
 \begin{equation}
     \d s^2 = -\left( \d t-g(z)\d z + \bar g (\bar z) \right)^2+f(z) \bar f(\bar z) \d z \d \bar z,
 \end{equation}
 where
 \begin{equation}
     f(z) = \prod_n (z-z_n)^{-4 G_N M_n},\hspace{2em} g(z) = 2i G_N\sum_n S_n \frac{\d z}{z- z_n}.
 \end{equation}
 % \begin{equation}
 %    \begin{aligned}
 %     \d s^2&=-\left(\d t-i4 G_N\sum_n\frac{S_n}{2}\left(\frac{\d z}{z-z_n}-\frac{\d\bar z}{\bar z-\bar z_n}\right)\right)^2+(z-z_n)^{-\frac{8 \pi G_N m_n}{2\pi}}(\bar z-\bar z_n)^{-\frac{8 \pi G_N m_n}{2\pi}}\d z\d \bar z\\
 %     &=-\left(\d t-g(z)\d z+\bar g(\bar z)\d \bar z\right)^2+f(z)\bar f(\bar z)\d z\d\bar z
 %    \end{aligned}
 % \end{equation}
 Now we can define the following vielbein corresponding to this metric:
 \begin{equation}
     e^a=\left(\begin{matrix}
         \d t-g(z)\d z+\bar g(\bar z)\d \bar z\\
         \ f( z) \d  z\\
         f(\bar z)\d \bar z
     \end{matrix}\right)^a.
 \end{equation}
 For $8 \pi G_N M_n/2\pi \notin\mathbb{Z}$ the functions $f(z)$ and $\bar f(\bar z)$ are multivalued, and so we restrict them with appropriate branch cuts $\gamma_n$. The vielbein can then be expressed in an explicitly Minkowski form by defining new coordinates
 \begin{equation}
     T(t,z,\bar z)=t-\int^z g(\zeta)d\zeta+\int^{\bar z}\bar g(\bar \zeta) d\bar \zeta\hspace{2em}Z(z)=\int^z f(\zeta)d\zeta\hspace{2em}\bar Z(\bar z)=\int^{\bar z} \bar f(\bar \zeta)d\bar\zeta
 \end{equation}
and identifying the edges on either side of the branch cut. In order for these new coordinates to be consistent we must choose an integration scheme which only follows contours in the holomorphic domain, and does not cross branch cuts. In these new coordinates, the spin connection is everywhere zero, and the vielbein is exact: $e^a = (\d T,\d Z,\d \bar Z)^a$. As in the case of the monopole, we see that going to explicitly Minkowski coordinates requires the introduction of non-trivial boundary conditions which encode the topological structure. We depict this situation in fig.~\ref{fig:placeholder}.

For the case of a single point mass, we were able to identify a gauge in which the full CSW gauge field is a solution to the abelian Chern--Simons equations of motion. This is also possible for the multicenter case, where to do this we Lorentz transform the vielbein to make its spatial components real and single-valued. Separating the magnitude and argument of the function $f(z)=R(z,\bar z) e^{i\Xi(z,\bar z)}$ with
\begin{equation}
    R(z,\bar z) = \prod_n[(z-z_n)(\bar z - \bar z_n)]^{-2 G_N M_n},\hspace{2em}\Xi(z,\bar z) =  -4 G_N\sum_n M_n \arg (z-z_n),
\end{equation}
we may define the Lorentz transformation
\begin{equation}
    \begin{aligned}
    \Lambda &= \exp \Big(-i\bm{J}_{12} \Xi(z,\bar z)\Big)\\
    &=\text{diag}\left[ 1, e^{ -i \Xi(z,\bar z)},e^{ i\Xi(z,\bar z)}\right].
    \end{aligned}
\end{equation}
The transformed vielbein will be given by
\begin{equation}
    \tilde e^a = \left(\begin{matrix}
        \d t - g(z) \d z +\bar g(\bar z)\d \bar z\\
        R(z,\bar z)\d z\\
        R(z,\bar z)\d z
    \end{matrix}\right)^a,
\end{equation}
and the spin connection by
\begin{equation}
    \frac{1}{2}\omega^{ab}\bm{J}_{ab} = \Lambda \d \Lambda^{-1} = -i\d \Xi(z,\bar z) \bm{J}_{12}.
\end{equation}
Now under a gauged translation parameterised by
\begin{equation}
    \begin{aligned}
        \tilde \rho^a &= {\Lambda^a}_b\left(\begin{matrix}t\\ Z(z)\\\bar Z(\bar z)\end{matrix}\right)^b=\left(\begin{matrix}
            t\\
            \exp[-i\Xi] \int^z f(\zeta)\d\zeta\\
            \exp[+i\Xi] \int^{\bar z} \bar f(\bar \zeta)\d\bar \zeta
        \end{matrix}\right)^a
    \end{aligned}
\end{equation}
the vielbein becomes
\begin{equation}
    \begin{aligned}
        \tilde e^a \rightarrow \tilde e^a - D\tilde \rho^a &= \tilde e^a - \left(\begin{matrix}
            \d t\\
            \left( -i \d \Xi(z,\bar z) Z(z) + f(z) \d z \right)e^{-i\Xi(z,\bar z)}\\
            \left(i \d\Xi(z,\bar z) \bar Z(\bar z)+\bar f(\bar z) \d\bar z\right)e^{i\Xi(z,\bar z)}
        \end{matrix}\right)-\left(\begin{matrix}
            0\\ i\d \Xi(z,\bar z) Z(z)e^{-i\Xi(z,\bar z)}\\ -i \d \Xi(z,\bar z) \bar Z(\bar z)e^{i\Xi(z,\bar z)}
        \end{matrix}\right)\\
        &=\left(\begin{matrix}
            -g(z)\d z+\bar g(\bar z) \d\bar z\\
            0\\0
        \end{matrix}\right).
    \end{aligned}
\end{equation}
Noting that the non-zero component of the vielbein is simply
\begin{equation}
    \bar g(\bar z)-g(z) = 2i G_N\,\d\left(\sum_n S_n \log\left(\frac{\bar z -\bar z_n}{z-z_n}\right)\right) = 4 G_N \,\d\left(\sum_n S_n \arg(z-z_n)\right),
\end{equation}
the full CSW gauge field solution can be written as
\begin{equation}
    \bm{A} = 4 G_N\d \left(\sum_n \left(m_n \bm{P}_0 +S_n\bm{J}_{12}\right) \arg(z-z_n)\right),
\end{equation}
which, as in the special case of the monopole, solves the abelian Chern--Simons equation of motion due to its Lie algebra value being in a Cartan subalgebra of the total algebra. Again, one may straightforwardly map this to a multicenter gauge theory solution, by replacing gravity generators by those in the Cartan subalgebra of a non-abelian gauge theory. To the best of our knowledge, this is the first time a multicenter classical solution has been mapped between gauge theory and gravity, and indeed it is interesting that this works. In a four-dimensional context, for example, single point charge / mass solutions in the Kerr-Schild approach of ref.~\cite{Monteiro:2014cda} cannot be straightforwardly extended to multicenter solutions without breaking the special (Kerr-Schild) properties of the solution. The fact that the relationship presented here -- that of reinterpreting colour generators in a gravity theory to obtain results in a doppelg\"{a}nger non-abelian gauge theory -- indeed corresponds to the conventional double copy is implicit in our general arguments presented earlier. For the benefit of the reader, however, we spell things out more explicitly in the following section.

\subsection{Doppelg\"angers and the conventional double copy}

The point-source spacetimes of \S\ref{sec:pointsource} are flat everywhere except at the particle locations, where the geometry carries distributional curvature and non-trivial holonomy. We now recast this same structure in more traditional double-copy language, using the Einstein--Hilbert formalism of gravity. To illustrate the point, it is sufficient to consider spinless particles, and our starting point is then the Einstein equations with an energy-momentum tensor corresponding to a system of point masses:
\[
G_{\mu\nu}=8\pi G_N\,T_{\mu\nu}
=8\pi G_N\sum_n M_n u_{n\mu}u_{n\nu}\,\delta^2(\mathbf{r}-\mathbf{r}_n).
\]
The corresponding line element can be obtained from eq.~(\ref{ds2multi}) (see also refs.~\cite{Deser:1983tn,Ortiz:1991gx}) as
\[
ds^2=-dt^2+\prod_n|\mathbf{r}-\mathbf{r}_n|^{-8G_NM_n}\big(dr^2+r^2d\theta^2\big).
\]
The Ricci curvature is
\[
R=16\pi G_N\sum_n M_n\,\delta^2(\mathbf{r}-\mathbf{r}_n),
\]
which vanishes away from $r=r_n$. A much stronger statement can be based on the fact that in $2+1$ dimensions, the full Riemann tensor is fixed by the Einstein tensor via
\[
R_{\mu\nu\rho\sigma}
=\varepsilon_{\mu\nu}{}^{\alpha}\varepsilon_{\rho\sigma}{}^{\beta}G_{\alpha\beta}
=8\pi G_N\,\varepsilon_{\mu\nu}{}^{\alpha}\varepsilon_{\rho\sigma}{}^{\beta}
\sum_n M_n u_{n\alpha}u_{n\beta}\,\delta^2(\mathbf{r}-\mathbf{r}_n).
\]
Thus, we see that there really is no curvature for $r\neq r_n$. This “flat-off-source” profile suggests that any gauge-theory avatar should also be flat away from sources and topological in character, matching the holonomy picture above. Chern–Simons theory with point charges provides exactly this. Take
\begin{equation}
\mathcal{L}_{\rm CS}
=\frac{k}{4\pi}\varepsilon^{\mu\nu\rho}\!\left(
A_\mu^a\partial_\nu A_\rho^a+\frac{1}{3}f^{abc}A_\mu^a A_\nu^b A_\rho^c\right) + A_\mu^a J^{\mu a},
\end{equation}
with equations of motion
  \begin{equation}
  \frac{k}{2\pi}\varepsilon^{\mu\nu\rho} F_{\nu\rho}^a=J^{\mu a},\qquad
  F_{\mu\nu}^a=\partial_\mu A_\nu^a-\partial_\nu A_\mu^a+f^{abc}A_\mu^b A_\nu^c,
  \label{eq:NAeom}
  \end{equation}
  and $D_\mu J^{\mu a}=0$. For static Wong charges with charge $g$, three-vector $u_n^\mu$ and colour $c_n^a$, we have
  \begin{equation}
  J^{\mu a}=g \sum_{n} u^\mu_n c_n^a,\delta^{(2)}(\vec r-\vec r_n),
  \end{equation}
  such that eq.~\eqref{eq:NAeom} gives
  \begin{equation}
  F^a_{\mu\nu}
  =\frac{2\pi g}{k}\varepsilon_{\mu\nu\rho}\sum_{n} u_n^\rho c^a_n,\delta^{(2)}(\vec r-\vec r_n)
  =\sum_n F_{n\,\mu\nu}^a,
  \label{eq:CSpointF}
  \end{equation}
  which is distributional and pure gauge for $r\neq r_n$, mirroring the gravitational case.

A well-established framework for writing down classical double copy relationships is the Weyl double copy of ref.~\cite{Luna:2018dpt}, which directly relates the Weyl tensor of (vacuum) gravity solutions in four spacetime dimensions to gauge theory field strengths\footnote{Reference~\cite{Luna:2018dpt} utilises the spinorial formalism of General Relativity. However, see ref.~\cite{Alawadhi:2020jrv} for a manifestly tensorial approach.}
This cannot be immediately generalised to $2+1$ dimensions, however, as the Weyl tensor vanishes identically. Nevertheless, we can express the above relationship between multicenter solutions through a double copy at the level of the Riemann tensor (or, relatedly, the Einstein tensor):
\begin{equation}
R_{\mu\nu\rho\sigma}
=\frac{k^2}{4\pi^2}\sum_n \frac{F^a_{n\,\mu\nu}F^a_{n\,\rho\sigma}}{\phi_n}
\Bigg|_{g c_n^a\rightarrow 8\pi G M_n u_n^a}.
\label{doublecopyR}
\end{equation}
Here, by analogy with the usual classical double copy, each $\phi_n$ corresponds to a biadjoint scalar field
\begin{equation}
    \phi_n^{a\bar{a}}=c_n^a c_n^{\bar{a}}\phi_n,
    \label{phiaabar}
\end{equation}
where $c^a$ and $c^{\bar{a}}$ are colour vectors. However, given the non-propagating nature of the gluon and graviton fields in 2+1 dimensions, the biadjoint fields in this case correspond to the Lagrangian and equation of motion
\[
\mathcal{L}=\phi^{a\bar a}\phi^{a\bar a}-\phi^{a\bar a}J^{a\bar a},\qquad
\phi^{a\bar a}=J^{a\bar a},
\]
describing non-propagating biadjoint modes. For point sources at locations $\vec{r}_i$, we can define the individual biadjoint field solutions
\[
\phi^{a\bar a}_n
=g c_n^a c_n^{\bar a}\,\delta^{(2)}(\vec r-\vec r_n),
\]
which then indeed lead to the double copy relationship of eq.~(\ref{doublecopyR}).

As reviewed in \S\ref{sec:review}, gravity in $2+1$ dimensions admits a Chern–Simons formulation. Writing
$$
S_{CSW}=\int d^3x~\eta_{ab}\!\left(\frac{k}{4\pi}\varepsilon^{\mu\nu\rho}e^a_\mu R^b_{\nu\rho}(\omega)
+ e_\mu^a T^{\mu b}\right),
$$
with $R_{\mu \nu}^a=\partial_\mu \omega_\nu^a-\pd_\nu \omega_\mu^a+\varepsilon^a{}_{bc}\omega_\mu^b\omega_\nu^c$ and
$$
T^{\mu a}=\sum_n M_n u_n^\mu u^a_n\,\delta^{(2)}(\vec r-\vec r_n),
$$
the equations of motion (after contracting with $\varepsilon_{\mu\nu\alpha}$) read
$$
R_{\nu\rho}^a
=8\pi G_N\,\varepsilon_{\mu\nu\alpha}T^{\alpha a}
=8\pi G_N\sum_n M_n \varepsilon_{\mu\nu\alpha}u_n^\alpha u^a_n\,\delta^{(2)}(\vec r-\vec r_n).
$$
In this guise, performing the same double copy as above amounts to a doppelg\"anger identification: instead of `squaring' the field strengths, one chooses the gauge algebra to be either $\mathfrak{g}$ (for the Yang–Mills Chern–Simons system) or $\mathfrak{iso}(2,1)$ (for the gravitational Chern–Simons system). The two descriptions are flat away from sources and encode identical holonomy data around each defect, matching the construction in \S\ref{sec:pointsource}.

\section{Conclusion}
\label{sec:conclude}

In this paper, we have examined the well-known double copy relationship between gauge theories and gravity, through the lens of alternative descriptions that may offer additional insights into both the origin and scope of this fascinating correspondence. We were motivated primarily by the question of what happens when one deliberately chooses formulations of gravity that look (almost) identical to non-abelian gauge theory. What does the double copy look like in such a framework? Can one gain useful insights into why the double copy works, or when it is expected to fail? To this end, we considered the useful playground of 2+1 dimensional gravity, which has a well-known description~\cite{Witten:1988hc} as a Chern--Simons theory whose gauge group is the Poincar\'{e} group ISO(2,1). In that approach, the double copy ceases to involve doubling at all, and becomes a precise doppelg\"{a}nger of non-abelian Chern--Simons (gauge) theory. We showed how known aspects of the double copy -- such as the relationship between Wilson lines in the two types of theory -- can be reproduced by taking the doppelg\"{a}nger approach as a starting point. Should one wish, one could then regard the latter as an origin of the classical double copy, that can in turn be used to test when the classical double copy is expected to break down.

To illustrate our analysis, we considered both single and multiple point masses / spins in gravity, and showed that in the CSW approach they could be written in a gauge such that only two commuting generators of the Poincar\'{e} group are turned on. These generators span a Cartan subalgebra of $\mathfrak{iso}(2,1)$, such that it is then straightforward to map the solutions to gauge theory counterparts, where one simply maps to appropriate Cartan subalgebra generators in the non-abelian gauge group. We showed that this map was equivalent to the traditional double copy, based on the Einstein--Hilbert formulation of 2+1 gravity. However, this approach is clearly not generally applicable. It may be the case, for example, that there is no gauge in which a CSW gravity solution can be written in terms of a closed subalgebra of the Poincar\'{e} group that can be mapped to a corresponding subalgebra of a compact gauge group. In such cases, it would not be possible to map explicit solutions between the theories, even though the equations of motion themselves map to each other. This potentially resolves a long-standing puzzle in the double copy literature, regarding whether it is possible to map {\it arbitrary} exact solutions, if equations of motion are known to double copy. It also sheds light on the fact that all exact position-space double copies that are currently known involve solutions that turn out to linearise their respective equations of motion. The same is true here, but we see a useful group theoretic justification behind this. 

Another question is whether the statements of this paper can be extended to higher dimensions. We have already seen a partial answer, in that we have uncovered an explicit double copy relationship between the Wong equations in Yang-Mills theory~\cite{Wong:1970fu}, and the MPD equations in gravity~\cite{Mathisson:1937zz,Papapetrou:1951pa,Dixon:1970zza}, that is dimension-independent. It is well-known, however, that precise formulations of gravity as a pure gauge theory do not generalise from 2+1 to many dimensions. Nevertheless, many alternative formulations of gravity exist~\cite{Krasnov:2020lku}, and it is highly likely that they can be used to further our understanding of the double copy. We thus hope that our paper is merely a first step in this direction, and look forward to further work in this area.

\section*{Acknowledgments}

This work has been supported by the UK Science and Technology
Facilities Council (STFC) Consolidated Grant ST/P000754/1 ``String
theory, gauge theory and duality''.

\bibliography{refs}
\end{document}